\PassOptionsToPackage{table,xcdraw}{xcolor}

\documentclass[sigconf]{acmart}

\AtBeginDocument{%
  \providecommand\BibTeX{{%
    \normalfont B\kern-0.5em{\scshape i\kern-0.25em b}\kern-0.8em\TeX}}}

\usepackage{makecell}
\usepackage{cleveref}
\usepackage{amsthm}
\usepackage{bm}
\usepackage[table,xcdraw]{xcolor}
\usepackage{multicol}
\usepackage{colortbl}
\usepackage{amsmath}
\usepackage{multirow}
\usepackage{subfigure}
\usepackage{enumitem}
\usepackage{booktabs}
\usepackage{bbding}
\usepackage{color}
\usepackage{xcolor}
\usepackage{booktabs}
\usepackage{hyperref}

\usepackage{algorithm}
\usepackage{listings}
\usepackage{natbib}

\usepackage[misc]{ifsym} 

\usepackage{etoolbox}
\makeatletter
\AfterEndEnvironment{algorithm}{\let\@algcomment\relax}
\AtEndEnvironment{algorithm}{\kern2pt\hrule\relax\vskip3pt\@algcomment}
\let\@algcomment\relax
\newcommand\algcomment[1]{\def\@algcomment{\footnotesize#1}}
\renewcommand\fs@ruled{\def\@fs@cfont{\bfseries}\let\@fs@capt\floatc@ruled
  \def\@fs@pre{\hrule height.8pt depth0pt \kern2pt}%
  \def\@fs@post{}%
  \def\@fs@mid{\kern2pt\hrule\kern2pt}%
  \let\@fs@iftopcapt\iftrue}
\makeatother

\acmSubmissionID{1218}

\AtBeginDocument{%
 \providecommand\BibTeX{{%
  Bib\TeX}}}

\copyrightyear{2023}
\acmYear{2023}
\setcopyright{rightsretained}
\acmConference[MM '23]{Proceedings of the 31st ACM International Conference on Multimedia}{October 29-November 3, 2023}{Ottawa, ON, Canada}
\acmBooktitle{Proceedings of the 31st ACM International Conference on Multimedia (MM '23), October 29-November 3, 2023, Ottawa, ON, Canada}
\acmDOI{10.1145/3581783.3611967}
\acmISBN{979-8-4007-0108-5/23/10}

\makeatletter
\gdef\@copyrightpermission{
  \begin{minipage}{0.3\columnwidth}
   \href{https://creativecommons.org/licenses/by/4.0/}{\includegraphics[width=0.90\textwidth]{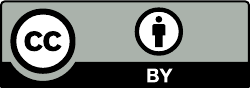}}
  \end{minipage}\hfill
  \begin{minipage}{0.7\columnwidth}
   \href{https://creativecommons.org/licenses/by/4.0/}{This work is licensed under a Creative Commons Attribution International 4.0 License.}
  \end{minipage}
  \vspace{5pt}
}
\makeatother

\begin{document}


\title[MISSRec: Pre-training and Transferring Multi-modal Interest-aware Sequence Representation \\ for Recommendation]{MISSRec: Pre-training and Transferring Multi-modal Interest-aware Sequence Representation for Recommendation}

\author{Jinpeng Wang}
\affiliation{%
  \institution{Tsinghua Shenzhen International Graduate School, Tsinghua University}
  \city{Shenzhen}
  \country{China}
}
\email{wjp20@mails.tsinghua.edu.cn}

\author{Ziyun Zeng}
\affiliation{%
  \institution{Tsinghua Shenzhen International Graduate School, Tsinghua University}
  \city{Shenzhen}
  \country{China}
}
\email{zengzy21@mails.tsinghua.edu.cn}

\author{Yunxiao Wang}
\affiliation{%
  \institution{Tsinghua Shenzhen International Graduate School, Tsinghua University}
  \city{Shenzhen}
  \country{China}
}
\email{wang-yx20@mails.tsinghua.edu.cn}

\author{Yuting Wang}
\affiliation{%
  \institution{Tsinghua Shenzhen International Graduate School, Tsinghua University}
  \city{Shenzhen}
  \country{China}
}
\email{wangyt22@mails.tsinghua.edu.cn}

\author{Xingyu Lu}
\affiliation{%
  \institution{Tsinghua Shenzhen International Graduate School, Tsinghua University}
  \city{Shenzhen}
  \country{China}
}
\email{luxy22@mails.tsinghua.edu.cn}

\author{Tianxiang Li}
\affiliation{%
  \institution{Tsinghua Shenzhen International Graduate School, Tsinghua University}
  \city{Shenzhen}
  \country{China}
}
\email{litx21@mails.tsinghua.edu.cn}

\author{Jun Yuan}
\affiliation{%
  \institution{Huawei Noah’s Ark Lab}
  \city{Shenzhen}
  \country{China}
}
\email{yuanjun25@huawei.com}


\author{Rui Zhang$^\text{\Letter}$}
\affiliation{%
  \country{www.ruizhang.info}
}
\email{rayteam@yeah.net}

\author{Hai-Tao Zheng}
\affiliation{%
  \institution{Tsinghua Shenzhen International Graduate School, Tsinghua University}
  \city{Shenzhen}
  \country{China}
}
\affiliation{%
  \institution{Peng Cheng Laboratory}
  \city{Shenzhen}
  \country{China}
}
\email{zheng.haitao@sz.tsinghua.edu.cn}

\author{Shu-Tao Xia$^\text{\Letter}$}
\affiliation{%
  \institution{Tsinghua Shenzhen International Graduate School, Tsinghua University}
  \city{Shenzhen}
  \country{China}
}
\affiliation{%
  \institution{Research Center of Artificial Intelligence, Peng Cheng Laboratory}
  \city{Shenzhen}
  \country{China}
}
\email{xiast@sz.tsinghua.edu.cn}

\thanks{\Letter\ Corresponding authors.}

\renewcommand{\shortauthors}{Jinpeng Wang et al.}

\newcommand{\dui}{\textcolor{teal}{\CheckmarkBold}}
\newcommand{\cuo}{\textcolor{purple}{\XSolidBrush}}
\newcommand{\ie}{\emph{i.e.},~}
\newcommand{\eg}{\emph{e.g.},~}
\newcommand{\wrt}{\emph{w.r.t.}~}
\newcommand{\aka}{\emph{aka}~}
\renewcommand{\paragraph}[1]{\medskip\noindent\textbf{#1.~}}
\newcommand{\todo}{\color{red}{\textbf{TODO.}}~}

\newcommand{\xiaok}[1]{\left(#1\right)}
\newcommand{\zhongk}[1]{\left[#1\right]}
\newcommand{\dak}[1]{\left\{#1\right\}}
\newcommand{\jiaok}[1]{\left<#1\right>}
\newcommand{\shuk}[1]{\left\lVert#1\right\rVert}
\newcommand{\shuks}[1]{\left\lVert#1\right\rVert^2}
\newcommand{\shangk}[1]{\left\lceil #1 \right\rceil}
\newcommand{\xiak}[1]{\left\lfloor #1 \right\rfloor}

\newcommand{\argmax}[1]{{\mathop{\arg\mathrm{max}}_{#1}\,}}
\newcommand{\argmin}[1]{{\mathop{\arg\mathrm{min}}_{#1}\,}}
\newcommand{\pfrac}[2]{\frac{\partial #1}{\partial #2}}
\newcommand{\prob}[2]{p\xiaok{#1 \mid #2}}

\newcommand{\T}{\top}
\newcommand{\dif}{\mathop{}\!\mathrm{d}}
\newcommand{\biset}[1]{\{0,1\}^{#1}}
\newcommand{\ReLU}{\mathrm{ReLU}}
\newcommand{\relu}{\mathrm{ReLU}}
\newcommand{\sign}{\mathrm{sign}}
\newcommand\softmax{\mathrm{softmax}}
\newcommand\KL{D_{\mathrm{KL}}}
\newcommand\Var{\mathrm{Var}}
\newcommand\Cov{\mathrm{Cov}}
\newcommand{\Tr}{\mathrm{Tr}}
\newcommand{\tr}{\mathrm{tr}}
\newcommand{\dist}{\mathrm{dist}}
\newcommand{\concat}{\mathrm{concat}}
\newcommand{\mean}{\mathrm{mean}}
\newcommand{\diag}{\mathrm{diag}}
\newcommand{\cov}{\mathrm{cov}}

\newcommand{\range}[1]{{0,1,\cdots,#1}} 
\newcommand{\Range}[1]{{1,2,\cdots,#1}} 
\newcommand{\opseq}[3]{{#1_1 #3 #1_2 #3 \cdots #3 #1_{#2}}}
\newcommand{\seq}[2]{\opseq{#1}{#2}{,}}
\newcommand{\xseq}[2]{\opseq{#1}{#2}{\times}}

\newcommand{\bma}{\bm{a}}
\newcommand{\bmb}{\bm{b}}
\newcommand{\bmc}{\bm{c}}
\newcommand{\bmd}{\bm{d}}
\newcommand{\bme}{\bm{e}}
\newcommand{\bmf}{\bm{f}}
\newcommand{\bmg}{\bm{g}}
\newcommand{\bmh}{\bm{h}}
\newcommand{\bmi}{\bm{i}}
\newcommand{\bmj}{\bm{j}}
\newcommand{\bmk}{\bm{k}}
\newcommand{\bml}{\bm{l}}
\newcommand{\bmm}{\bm{m}}
\newcommand{\bmn}{\bm{n}}
\newcommand{\bmo}{\bm{o}}
\newcommand{\bmp}{\bm{p}}
\newcommand{\bmq}{\bm{q}}
\newcommand{\bmr}{\bm{r}}
\newcommand{\bms}{\bm{s}}
\newcommand{\bmt}{\bm{t}}
\newcommand{\bmu}{\bm{u}}
\newcommand{\bmv}{\bm{v}}
\newcommand{\bmw}{\bm{w}}
\newcommand{\bmx}{\bm{x}}
\newcommand{\bmy}{\bm{y}}
\newcommand{\bmz}{\bm{z}}
\newcommand{\bmzero}{\bm{0}}
\newcommand{\bmone}{\bm{1}}
\newcommand{\bmalpha}{\bm{\alpha}}
\newcommand{\bmbeta}{\bm{\beta}}
\newcommand{\bmgamma}{\bm{\gamma}}
\newcommand{\bmdelta}{\bm{\delta}}
\newcommand{\bmepsilon}{\bm{\epsilon}}
\newcommand{\bmtheta}{\bm{\theta}}
\newcommand{\bmiota}{\bm{\iota}}
\newcommand{\bmkappa}{\bm{\kappa}}
\newcommand{\bmlambda}{\bm{\lambda}}
\newcommand{\bmmu}{\bm{\mu}}
\newcommand{\bmnu}{\bm{\nu}}
\newcommand{\bmxi}{\bm{\xi}}
\newcommand{\bmpi}{\bm{\pi}}
\newcommand{\bmrho}{\bm{\rho}}
\newcommand{\bmsigma}{\bm{\sigma}}
\newcommand{\bmtau}{\bm{\tau}}
\newcommand{\bmupsilon}{\bm{\upsilon}}
\newcommand{\bmphi}{\bm{\phi}}
\newcommand{\bmchi}{\bm{\chi}}
\newcommand{\bmpsi}{\bm{\psi}}
\newcommand{\bmomega}{\bm{\omega}}
\newcommand{\bmA}{\bm{A}}
\newcommand{\bmB}{\bm{B}}
\newcommand{\bmC}{\bm{C}}
\newcommand{\bmD}{\bm{D}}
\newcommand{\bmE}{\bm{E}}
\newcommand{\bmF}{\bm{F}}
\newcommand{\bmG}{\bm{G}}
\newcommand{\bmH}{\bm{H}}
\newcommand{\bmI}{\bm{I}}
\newcommand{\bmJ}{\bm{J}}
\newcommand{\bmK}{\bm{K}}
\newcommand{\bmL}{\bm{L}}
\newcommand{\bmM}{\bm{M}}
\newcommand{\bmN}{\bm{N}}
\newcommand{\bmO}{\bm{O}}
\newcommand{\bmP}{\bm{P}}
\newcommand{\bmQ}{\bm{Q}}
\newcommand{\bmR}{\bm{R}}
\newcommand{\bmS}{\bm{S}}
\newcommand{\bmT}{\bm{T}}
\newcommand{\bmU}{\bm{U}}
\newcommand{\bmV}{\bm{V}}
\newcommand{\bmW}{\bm{W}}
\newcommand{\bmX}{\bm{X}}
\newcommand{\bmY}{\bm{Y}}
\newcommand{\bmZ}{\bm{Z}}
\newcommand{\bmGamma}{\bm{\Gamma}}
\newcommand{\bmDelta}{\bm{\Delta}}
\newcommand{\bmTheta}{\bm{\Theta}}
\newcommand{\bmLambda}{\bm{\Lambda}}
\newcommand{\bmXi}{\bm{\Xi}}
\newcommand{\bmPi}{\bm{\Pi}}
\newcommand{\bmSigma}{\bm{\Sigma}}
\newcommand{\bmUpsilon}{\bm{\Upsilon}}
\newcommand{\bmPhi}{\bm{\Phi}}
\newcommand{\bmPsi}{\bm{\Psi}}
\newcommand{\bmOmega}{\bm{\Omega}}

\newcommand{\calA}{\mathcal{A}}
\newcommand{\calB}{\mathcal{B}}
\newcommand{\calC}{\mathcal{C}}
\newcommand{\calD}{\mathcal{D}}
\newcommand{\calE}{\mathcal{E}}
\newcommand{\calF}{\mathcal{F}}
\newcommand{\calG}{\mathcal{G}}
\newcommand{\calH}{\mathcal{H}}
\newcommand{\calI}{\mathcal{I}}
\newcommand{\calJ}{\mathcal{J}}
\newcommand{\calK}{\mathcal{K}}
\newcommand{\calL}{\mathcal{L}}
\newcommand{\calM}{\mathcal{M}}
\newcommand{\calN}{\mathcal{N}}
\newcommand{\calO}{\mathcal{O}}
\newcommand{\calP}{\mathcal{P}}
\newcommand{\calQ}{\mathcal{Q}}
\newcommand{\calR}{\mathcal{R}}
\newcommand{\calS}{\mathcal{S}}
\newcommand{\calT}{\mathcal{T}}
\newcommand{\calU}{\mathcal{U}}
\newcommand{\calV}{\mathcal{V}}
\newcommand{\calW}{\mathcal{W}}
\newcommand{\calX}{\mathcal{X}}
\newcommand{\calY}{\mathcal{Y}}
\newcommand{\calZ}{\mathcal{Z}}

\newcommand{\bbC}{\mathbb{C}}
\newcommand{\bbE}{\mathbb{E}}
\newcommand{\bbI}{\mathbb{I}}
\newcommand{\bbN}{\mathbb{N}}
\newcommand{\bbQ}{\mathbb{Q}}
\newcommand{\bbR}{\mathbb{R}}
\newcommand{\bbZ}{\mathbb{Z}}

\newcommand{\tabincell}[2]{\begin{tabular}{@{}#1@{}}#2\end{tabular}}

\newcommand{\modelname}{\textsf{MISSRec}}
\begin{abstract}
The goal of sequential recommendation (SR) is to predict a user's potential interested items based on her/his historical interaction sequences. 
Most existing sequential recommenders are developed based on ID features, which, despite their widespread use, often underperform with sparse IDs and struggle with the cold-start problem. 
Besides, inconsistent ID mappings hinder the model's transferability, isolating similar recommendation domains that could have been co-optimized. 
This paper aims to address these issues by exploring the potential of multi-modal information in learning robust and generalizable sequence representations. 
We propose \textbf{\modelname{}}, a multi-modal pre-training and transfer learning framework for SR. 
On the user side, we design a Transformer-based encoder-decoder model, where the contextual encoder learns to capture the 
sequence-level multi-modal user interests 
while a novel interest-aware decoder is developed to grasp item-modality-interest relations for better sequence representation. 
On the candidate item side, we adopt a dynamic fusion module to produce user-adaptive item representation, providing more precise matching between users and items. 
We pre-train the model with contrastive learning objectives and fine-tune it in an efficient manner. 
Extensive experiments demonstrate the effectiveness and flexibility of \modelname{}, promising a practical solution for real-world recommendation scenarios. 
Data and code are available on \url{https://github.com/gimpong/MM23-MISSRec}.
\end{abstract}

%
%

%
%

\ccsdesc[500]{Information systems~Recommender systems}
\ccsdesc[500]{Information systems~Multimedia information systems}
\ccsdesc[500]{Information systems~Personalization}

%
\keywords{multi-modal sequential recommendation, pre-training, parameter-efficient fine-tuning, interest-aware sequence representation}

\maketitle

\section{Introduction}
\label{sec: introduction}

\begin{figure}[t]
    \centering
    \includegraphics[width=\columnwidth]{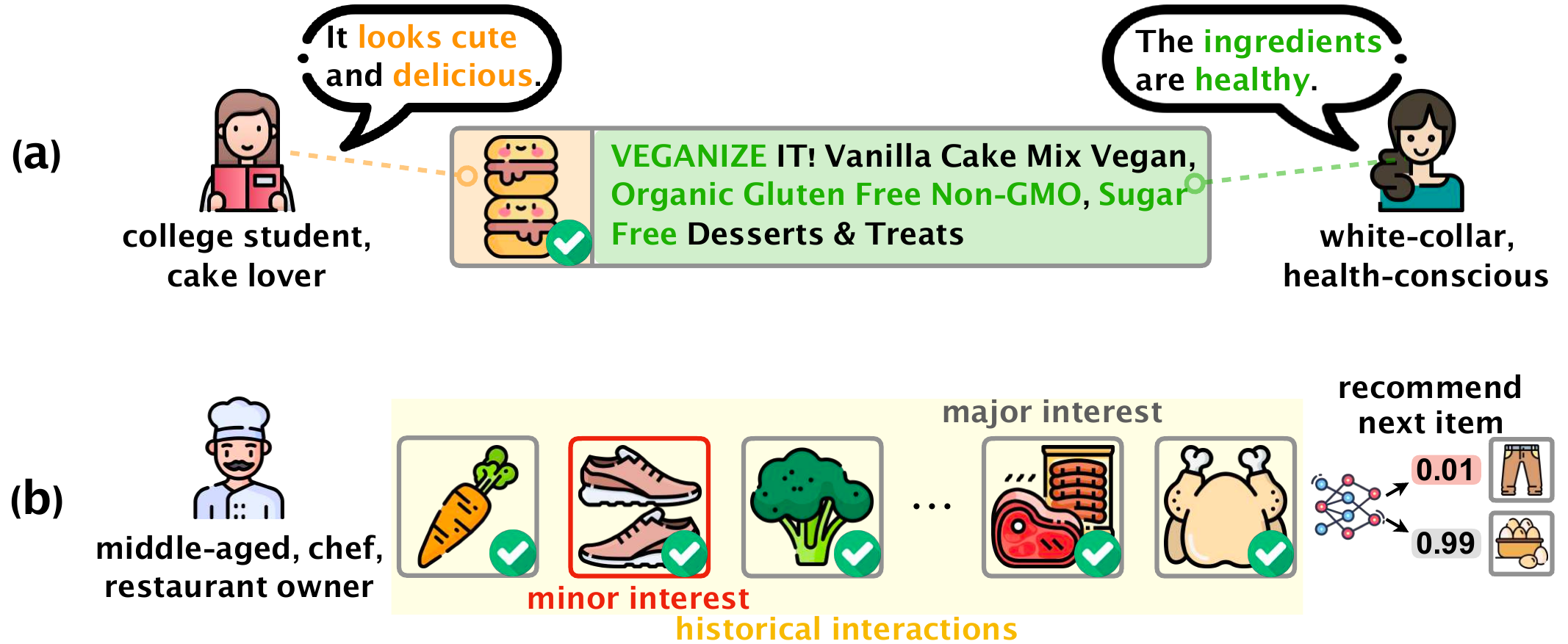}
    \caption{Challenges of multi-modal sequential recommendation. 
    (a) The multi-modal synergy within each item can be user-dependent and dynamic. 
    Users may interact with the same item focusing on different modalities. 
    (b) Redundant interaction may overwhelm sparse but important interests. Modeling interactions equally will lead to biased predictions. 
    }
    \label{fig:intro}
\end{figure}

Recommendation systems~\cite{RecSys,LightGCN,su2021detecting,su2022detecting} are an important component in various application domains, from e-commerce to content platforms, improving user engagement and satisfaction. 
Sequential recommendation (SR)~\cite{SR_survey1,SR_survey2}, a popular methodology of recommendation, aims to predict users' potential interested items according to their historical interaction sequences. 
Typically, this task is formulated as a representation learning problem, where user embeddings are learned to capture underlying preferences behind the interacted item sequence and to match with suitable item embeddings. 

Current SR approaches~\cite{SASRec,BERT4Rec,S3-Rec,ICLRec,DFAR} predominantly rely on ID features as input, which, despite their widespread uses in practice, exhibit two primary shortcomings. 
(\textbf{i}) The robustness and scalability largely depend on the distribution of user-item interaction. 
Due to the sparsity of interaction data, it is challenging to learn accurate item and sequence representations based on ID information, let alone the cold-start problem~\cite{coldstart_problem} for new items. 
Meanwhile, ID-based approaches often exhibit a popularity bias toward popular IDs~\cite{PopBias1,PopBias2}, posing the fairness issue for the majority of infrequent IDs. 
(\textbf{ii}) Transferring knowledge to new scenarios is hard because of inconsistent ID mappings. It isolates similar domains that can be co-optimized, limiting the applicability of ID-based models. 

Given the ubiquity of text and images in real-world scenarios, leveraging multi-modal data is promising to remedy the above shortcomings~\cite{yang2021unimf}. 
Particularly, both text and visual information play crucial roles in attracting user attention -- compelling titles can boost user engagement, while visual elements like colors and shapes, can influence user decisions. 
Considering remarkable progress in computer vision~\cite{ViT_Survey1,ViT_Survey2} and natural language understanding~\cite{Prompt_survey,LLM_survey}, they offer robust and generalizable semantic extraction capability to item content. 
Therefore, we believe it necessary and also feasible to harness multi-modal content understanding to the SR. 
However, such investigation remains in a fledgling stage, as our literature review (\S\ref{sec:related_work}) will show. 
\emph{How does multi-modal information impact SR? 
How to make better use of such information in sequence modeling?}
These two questions drive our study in this paper. 

In general, utilizing multi-modal information in SR poses 
two challenges. 
(\textbf{i}) The multi-modal synergy within each item is user-dependent and dynamic. 
Users may interact with the same item for various reasons, and different modalities contribute unequally to user interest in an item. 
Even for the same user-item pair, such patterns can be time- and context-varying, making it difficult to design effective multi-modal fusion. 
(\textbf{ii}) Information redundancy can overwhelm essential user interests. 
Interaction sequences typically exhibit imbalanced interest distribution. 
A user's sequence usually contains lots of homogeneous items, \eg daily necessities, while other informative items, \eg sports goods, may sparsely appear. 
If we treat all interactions equally in user behavior modeling, it will result in overemphasis on specific kinds of items and insufficient focus on others. 
We give an illustration of these challenges in \cref{fig:intro}. 

To tackle these challenges, we propose \textbf{\modelname{}} (\underline{M}ulti-modal \underline{I}ntere\underline{S}t-aware \underline{S}equential \underline{Rec}ommendation), a multi-modal pre-training and transferring framework for SR. 
In general, \modelname{} takes interaction sequences with multi-modal item information as input, learns ID-agnostic representations with an interest-aware encoder-decoder model via self-supervised pre-training, and can be efficiently adapted for multiple domains to enhance recommendation. 
Specifically, (\textbf{i}) to bridge the semantic gap between general domains and recommendation, we incorporate multi-modal feature adapters~\cite{Adapter1,Adapter2} to distill personalized semantics of general multi-modal features in a parameter-efficient manner. 
(\textbf{ii}) To explore the multi-modal synergy in the user-item interaction process, 
we 
introduce a lightweight dynamic fusion strategy to produce user-specific item representations. 
(\textbf{iii}) To mitigate the adverse effect of information redundancy on sequence modeling, we introduce an interest discovery module to mine global multi-modal interests among users. 
By associating items with relevant interests via adaptive clustering, we convert multi-modal sequences into a series of interest tokens. 
(\textbf{iv}) We design a novel Transformer-based encoder-decoder architecture for interest-aware multi-modal sequence modeling, where the contextual encoder captures the relations among multi-modal user interests and the decoder grasps essential item-modality-interest patterns for comprehensive sequence representation. 

We conduct detailed experiments and show the effectiveness of \modelname{} in leveraging multi-modal data to learn robust and transferable sequence representation. 
In particular, it exhibits better generalizability than ID-based and text-based approaches and further alleviates the item cold-start issue. 

To summarize, we make the following contributions in this paper.
\setlist{nolistsep}
\begin{itemize}[leftmargin=1.5em]
\item We highlight the significance and challenges of exploiting multi-modal information for SR and propose an effective pre-training and efficient transfer learning framework for it. 
\item To capture the dynamic multi-modal synergy in the user-item interaction process, we 
adopt a lightweight dynamic fusion module to produce user-adaptive candidate item representation. 
\item We introduce a multi-modal interest discovery module, based on which we construct an interest-aware decoder to model item-modality-interest relations for better sequence representation. 
\item Extensive experiments of pre-training and downstream domain adaptation justify the comprehensive merits of our approach. 
\end{itemize}

\section{Related Works}
\label{sec:related_work}
\subsection{Sequential Recommendation}
Sequential recommendation (SR) aims to predict the next suitable item for a given user based on her/his interacted item sequence~\cite{SR_survey1,SR_survey2}. 
Early SR approaches~\cite{Fossil,FPMC} exploited the Markov chain mechanism for sequential modeling, while deep SR approaches explored CNN-~\cite{Caser,NextINet}, RNN-~\cite{DUPN,GRU4Rec}, MLP-~\cite{FMLP-Rec,MMMLP} and attention-based~\cite{SASRec,HGN,FDSA,BERT4Rec} architectures to model the transition patterns in the interaction sequences. 
Besides, interest modeling~\cite{DIN,MIND,ComiRec,DFAR,Re4} has been another popular methodology for SR, where user interests were usually implemented by attention or clustering. 
From other perspectives, extensive efforts have been devoted to designing effective learning strategies for SR, including temporal-aware learning tasks~\cite{TiSASRec,TCPSRec,RESETBERT} and contrastive learning objectives~\cite{CL4SRec,ICLRec,CL4SRec,S3-Rec,DuoRec,ContraRec}. 
Note that most existing SR approaches are designed with ID features, \eg item IDs or attribute IDs. 
They usually suffer from the cold-start problem and fall short of transferability. 
Several SR approaches have leveraged multi-modal item contents to mitigate these issues. 
For instance, for each item, CSAN~\cite{CSAN} aggregated multiple information including multi-modal features and projected the result to obtain an embedding, providing better item representation than the ID embedding alone. 
MM-Rec~\cite{MM-Rec} adopted pre-trained VL-BERT~\cite{VL-BERT} as the multi-modal item encoder and built a single-tower model for SR. 
Inspired by the success of MLP-Mixer~\cite{MLP-Mixer} in CV, MMMLP~\cite{MMMLP} adapted the architecture to multi-modal SR. 
On the basis of these works, we further focus on learning universal sequence representations with multi-modal item content. 
Our \modelname{} contributes a pre-training and efficient transfer framework for multi-modal SR, which can benefit real-world practice. 

\subsection{Pre-training and Transfer Learning in Recommendation}
The paradigm of ``pre-train and transfer'' has become increasingly popular in recommendation. 
Compared to cross-domain recommendation~\cite{CDR_Survey}, this paradigm is more general as it does not require cross-domain correspondence, \eg overlapped items. 
CLUE~\cite{CLUE} and PeterRec~\cite{PETERec} designed ID-based sequence models and adopted contrastive learning objectives~\cite{MoCo,SimCLR} for model learning, where PeterRec further made the subsequent transfer parameter-efficient. 
UPRec~\cite{UPRec} leveraged user profiles and social information to construct auxiliary pre-text tasks. 
The assumption that both pre-training and target domains share ID vocabulary restricts their application scopes. 
By contrast, modality-based methods are more flexible. 
For instance, text-based SR pre-training  approaches~\cite{UniSRec,VQRec} utilized a pre-trained language model (PLM, \eg BERT~\cite{BERT}) as the frozen feature extractor in item representation modules, showing efficacy and efficiency. 
\citet{MoRec} designed text-based and image-based SR models under different scenarios, demonstrating the improvement from the tunable feature extractors. 
\citet{TransRec} considered mixed modality sequences where each item is either a text or an image. They also chose to optimize the SR model and feature extractors jointly, improving the item representation but largely downgrading learning efficiency. 
Furthermore, to capture the multi-modal synergy within each item, two latest works~\cite{MSM4SR,MMSRec} have focused on SR pre-training and transfer learning with multi-modal item representations. 
Our \modelname{} also belongs to this setting, but we have fulfilled 
two 
improvements. 
(\textbf{i}) To our best knowledge, we are the first to design interest-aware modeling for multi-modal SR. 
(\textbf{ii}) We design user-adaptive candidate item fusion to model users' dynamic attention to different modalities.

\subsection{Multi-modal Recommendation}
Leveraging multi-modal information (\eg text and images) has been shown to improve the recommendation efficacy in many applications, \eg news feed~\cite{UNBERT,IMRec,liu2022boosting,MINER}, short-video feed~\cite{SEMI,jiang2020aspect,cai2021heterogeneous}, fashion e-commerce~\cite{VBPR,song2023mm}, and can effectively alleviate cold-start issues~\cite{MML,MTPR}. 
Therefore, multi-modal recommendation (MMR) has become an active research topic~\cite{MMRec_Survey1,MMRec_Survey2,MMRec_Survey3}. 
Early approaches explored MMR with matrix factorization models~\cite{VBPR,ACF}, while the latter ones mainly developed GNN-based models~\cite{MMGCN,GRCN,MKGAT} to demonstrate the efficacy. 
On this basis, recent works have delved into self-supervised learning strategies~\cite{BM3,LATTICE,SLMRec,MMSSL,MMGCL,LCD} to enhance in-domain robustness or pre-training strategies~\cite{U-BERT,PMGT,MMCPR,CHEST,CPT-HG,HyperCTR} to improve cross-domain generalization. 
Despite the remarkable progress in MMR, existing solutions were predominantly designed for non-sequential scenarios, \eg collaborative filtering~\cite{SimpleX}, which are not very suitable for SR. 
Fortunately, our work uncovers some challenges in multi-modal SR (see \S\ref{sec: introduction} and \cref{fig:intro}) and fulfills effective solutions. We hope it can provide a research basis for future work. 
\begin{figure*}[t]
    \centering
    \includegraphics[width=\textwidth]{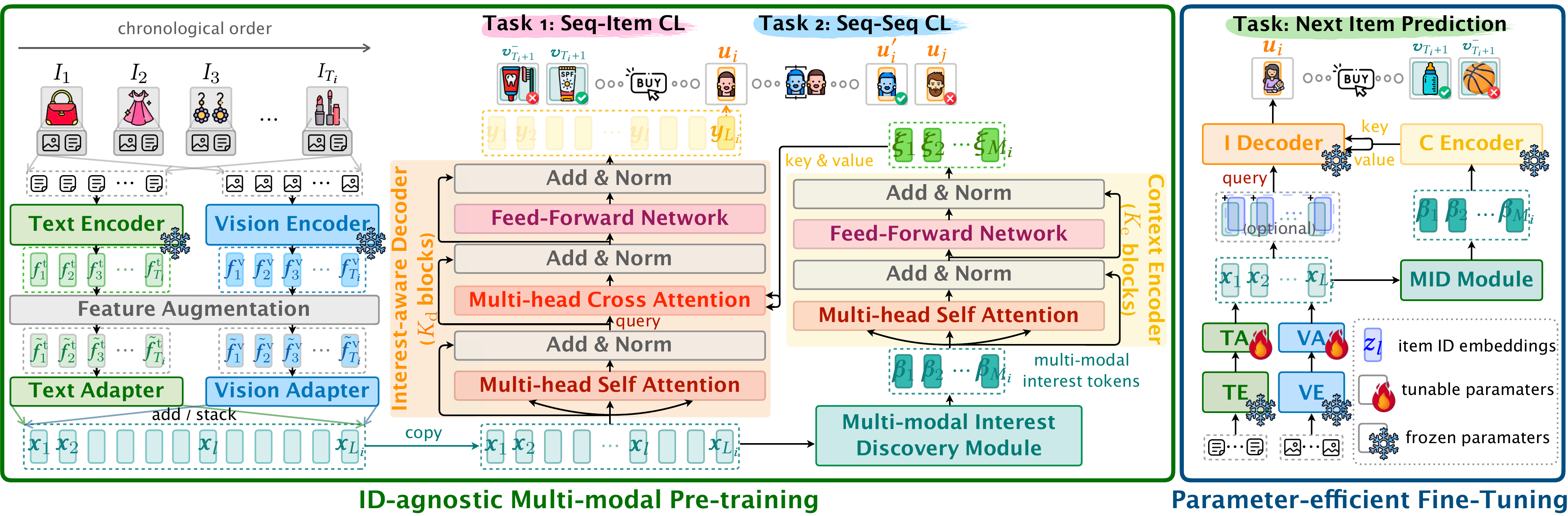}
    \vspace{-2em}
    \caption{\modelname{} consists of two stages. 
    In the first stage, it extracts multi-modal features from item content and transforms them into a token sequence. 
    Meanwhile, it adaptively converts the token sequence into multi-modal interest tokens. 
    Then, it encodes the mined interest sequence to capture contextual cues for personalization. 
    Next, by developing an interest-aware decoding mechanism, it produces a comprehensive sequence embedding. 
    Finally, it adopts sequence-item and sequence-sequence contrastive objectives for pre-training. 
    In the second stage, it uses the sequence-item objective for fine-tuning. 
    Both transductive (\ie w/ item IDs) and inductive (\ie ID-agnostic) tuning settings are supported.  
    The downstream domain adaptation is parameter-efficient as only the modality-specific adapters need to be tuned. 
    \emph{This figure is best viewed in color.}}
    \label{fig:arc}
\end{figure*}

\section{Method}
\label{sec:method}

\subsection{Problem Formulation and Method Overview}
\label{subsec:overview}
Given the historical interaction sequence of the $i$-th user, $S_{i}=[ I_1, I_2, \cdots, I_{T_i} ]$, ordered by timestamps, sequential recommendation (SR) aims to predict the next item $I_{T_i+1}$ that the user may interact with by learning representations for $S_{i}$ and candidate items. 
Under the multi-modal settings, each item contains a unique item ID and multi-modal content. 
Without loss of generality, we study leveraging text and image modalities to improve SR in this paper. 
We propose \textbf{\modelname{}}, a pre-training and transferring framework, as illustrated in \cref{fig:arc}. 
\modelname{} consists of 7 main components:
(\textbf{i}, \S\ref{subsubsec:feat_extr}) Pre-trained, frozen text and image encoders for extracting multi-modal features. 
(\textbf{ii}, \S\ref{subsubsec:adapter}) A pair of modality-specific adapters to transform multi-modal features into input tokens, which bridge the semantic gap between general features and personalization. 
In the fine-tuning stage, they help with efficient model adaptation. 
(\textbf{iii}, \S\ref{subsubsec:candidate_item_enc}) A dynamic fusion module for generating user-adaptive candidate item representations. 
(\textbf{iv}, \S\ref{subsubsec:cluster_interests}) A multi-modal interest discovery module that generates user interest tokens for the decoder. 
(\textbf{v}, \S\ref{subsubsec:mm_enc}) A contextual Transformer encoder for the interest sequence that dynamically grasps important cues for personalization.
(\textbf{vi}, \S\ref{subsubsec:mm_dec}) An interest-aware Transformer decoder to capture the item-modality-interest relation for better sequence representation. 
(\textbf{vii}, \S\ref{subsec:pretrain} and \S\ref{subsec:fine_tuning}) The pre-training and downstream fine-tuning tasks to achieve transferable recommendation. 

\subsection{Universal Multi-modal Item Representation}
\label{subsec:item_rep} 
\subsubsection{Multi-modal Feature Extraction with Pre-trained Transformers}
\label{subsubsec:feat_extr}
\modelname{} utilizes multi-modal features to achieve universal representations. 
Concretely, we take advantage of the pre-trained transformers for item content understanding. 
Specifically, we use BERT~\cite{BERT} and ViT~\cite{ViT} for text and visual feature extraction, respectively. 
We adopt the cross-modal pre-trained version~\cite{CLIP} following \citet{MSM4SR}. 
For an item $I$, we obtain its text feature vector by $\bmf^\text{t}=\phi^\text{t}(I)$ and the image feature vector by $\bmf^\text{v}=\phi^\text{v}(I)$, where $\phi^\text{t}$ and $\phi^\text{v}$ denotes the text encoder the visual encoder, respectively. 

\subsubsection{Feature Augmentation}
\label{subsubsec:feat_aug}
To enhance the robustness of the sequence representation model, we apply dropout~\cite{Dropout} as the feature augmentation. 
Given the text features $\bmF^\text{t}_i=[\seq{\bmf^\text{t}}{T_i}]$, we apply twice the feature augmentation and subsequently obtain $\tilde{\bmF}_i^\text{t}=[\seq{\tilde{\bmf}^\text{t}}{T_i}]$ and 
$\tilde{\bmF}_i^{\text{t}'}=[\seq{\tilde{\bmf}^{\text{t}'}}{T_i}]$ as the augmented text features. 
They serve as two positive views \wrt $\bmF^\text{t}_i$. 
The image features $\bmF^\text{v}_i$ follow analogous augmentation procedure.

\subsubsection{Bridge Domain Gap with Modality-specific Adapters}
\label{subsubsec:adapter}
We adopt modality-specific adapters~\cite{Adapter1,Adapter2} to reduce irrelevant semantics in multi-modal features and enhance the personalization. 
We transform the augmented features and fuse the results as item tokens, for example, $\bmX_i=[\seq{\bmx}{L_i}]=[\psi^\text{t}(\tilde{F}^\text{t}_i);\psi^\text{v}(\tilde{F}^\text{v}_i)]$. $\psi^\text{t}$ and $\psi^\text{v}$ denote the text and visual adapters, respectively.

In contrast to tunable feature extractors, adapters with frozen extractors exhibits two strengths. 
(\textbf{i}) It allows (pre-) training of the sequence model based on pre-extracted features, reducing time and memory overhead. 
(\textbf{ii}) Adapters enable parameter-efficient transfer in downstream domains, which is preferable to fine-tuning.  

\subsubsection{Candidate Item Representation with Dynamic Fusion} 
\label{subsubsec:candidate_item_enc}
To model users' dynamic attention to different modalities of candidate items, we design a lightweight fusion module to generate user-adaptive item representations. 
Given a candidate item $I_k$, we first encode it with modality encoders and adapters to obtain modality embeddings, \ie $\bmx_k^\text{t}$ and $\bmx_k^\text{v}$. 
For the $i$-th user, we compute its sequence representation $\bmu_i$ according to the procedures in \S\ref{subsec:mid_module} and \S\ref{subsec:mm_enc_dec}.
Then, the adaptive representation of $I_k$ is defined by the weighted fusion of modality embeddings, namely
\begin{gather}\label{equ:user-adaptive-item-rep}
    \bmv_k = 
    \frac{
    e^{\alpha\cdot\langle \bmu_i,\bmx_k^\text{t} \rangle}\cdot\bmx_k^\text{t}+
    e^{\alpha\cdot\langle \bmu_i,\bmx_k^\text{v} \rangle}\cdot\bmx_k^\text{v}
    }{
    e^{\alpha\cdot\langle \bmu_i,\bmx_k^\text{t} \rangle}+
    e^{\alpha\cdot\langle \bmu_i,\bmx_k^\text{v} \rangle}
    }.
\end{gather}
$\langle\cdot,\cdot\rangle$ denotes the inner product.
$\alpha\ge0$ is a concentration factor to balance two modalities. 
Besides the efficacy, the fusion keeps efficiency as well, because it is factorizable when computing matching scores. 
Specifically, let us define $s^\text{t}_{i,k}=\langle\bmu_i,\bmv_j^\text{t}\rangle$ and $s^\text{v}_{i,k}=\langle\bmu_i,\bmv_j^\text{v}\rangle$, the matching score between $\bmu_i$ and $\bmv_k$ can be compute by
\begin{gather}\label{equ:factorizable-ui-match-score}
    \langle\bmu_i,\bmv_k\rangle=\frac{s^\text{t}_{i,k}\cdot e^{\alpha\cdot s^\text{t}_{i,k}}+s^\text{v}_{i,k}\cdot e^{\alpha\cdot s^\text{v}_{i,k}}
    }
    {e^{\alpha\cdot s^\text{t}_{i,k}}+e^{\alpha\cdot s^\text{v}_{i,k}}}.
\end{gather}
We can equivalently pre-compute modality-specific scores and fuse them to get the overall score. 
In particular, dynamic fusion exhibits a transition between mean and max poolings: 
$\langle\bmu_i,\bmv_k\rangle=(s^\text{t}_{i,k}+s^\text{v}_{i,k})/2$ when $\alpha=0$, while $\langle\bmu_i,\bmv_k\rangle=\max(s^\text{t}_{i,k},s^\text{v}_{i,k})$ when $\alpha\rightarrow+\infty$. 

\begin{figure}[t]
    \centering
    \includegraphics[width=\columnwidth]{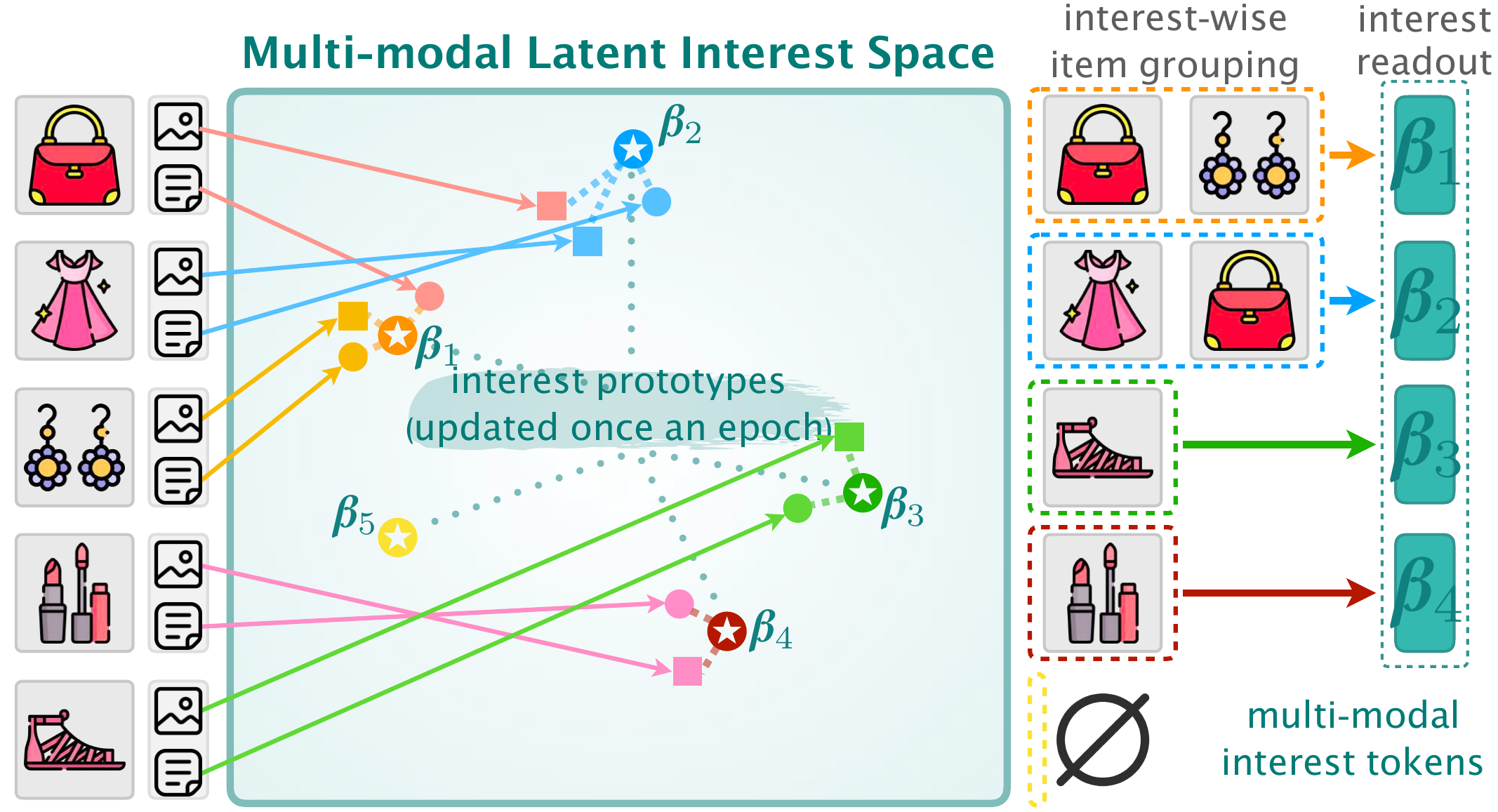}
    \vspace{-2em}
    \caption{Convert item tokens to interest tokens via the multi-modal interest discovery (MID) module. An item can be converted to different interests according to multiple modalities.}
    \label{fig:mid_module}
\end{figure}

\subsection{Discover Multi-modal User Interests}
\label{subsec:mid_module}
Typically, item sequences reflect users' implicit interests that are important for SR. 
Interest modeling can reduce the negative impact of information redundancy, helping to capture personalized semantics for precise representation. 
In this sub-section, we introduce a multi-modal interest discovery (MID) module to assist this goal. 

\subsubsection{Excavate and Index Interests via Adaptive Clustering}
\label{subsubsec:cluster_interests}
During the training process, item tokens are expected to convey both transferable multi-modal information and personalized semantics. 
Hence, we adopt a variant of $k$-nearest neighbor-based density peaks clustering algorithm (DPC-KNN)~\cite{DPC-KNN} on the whole item token set $\calX=\{\bmx_i\}_{i=1}^{N_I}$, to explore the distribution of user interests before every training epoch starts. 
First, for the $i$-th token, we compute its local density by considering its $k$-nearest neighbors:
\begin{gather}
    \rho_i=\exp\xiaok{-\frac{1}{k}\sum_{\bmx_j\in k\text{NN}(\bmx_i)}\shuk{\bmx_i-\bmx_j}_2^2}.
\end{gather}
Then, for the $i$-th token, we compute its minimum distance to any other token with higher local density, namely 
\begin{gather}
    \delta_i=
    \begin{cases}
        \min_{j:\rho_j>\rho_i}\shuk{\bmx_i-\bmx_j}_2, & \text{if }\exists j\text{ s.t. }\rho_j>\rho_i, \\
        \max_{j}\shuk{\bmx_i-\bmx_j}_2, &\text{otherwise}.
    \end{cases}
\end{gather}
Next, we select $K_\text{c}$ cluster centroids with the highest $\rho_i\times\delta_i$ scores as the interest prototypes. 
Finally, we index each token by assigning it to the nearest centroid (\ie the interest prototype). 

\subsubsection{Translate Multi-modal Item Tokens into Interest Tokens}
\label{subsubsec:transform_interest}
As shown in \cref{fig:arc,fig:mid_module}, the MID module receives item tokens, finds the nearest interest prototype for each token, and returns irredundant interest tokens. 
To accelerate token conversion, we 
can build
the interest index of each item token after clustering for fast lookup. 

\subsection{Multi-modal Interest-aware Sequence Model}
\label{subsec:mm_enc_dec}
We design an encoder-decoder model for sequence representation, which unifies sequence modeling and multi-modal fusion. 
\subsubsection{Transformer-based Multi-modal Context Encoder}
\label{subsubsec:mm_enc}
To pursue personalization, we expect the model to adaptively capture crucial cues from the given context, \ie discovered interest tokens (\S\ref{subsubsec:transform_interest}), denoted by $\seq{\bmbeta}{M_i}$. 
We construct an encoder with $K_\text{e}$ Transformer encoder blocks, each of which follows the standard design~\cite{attention}. 
The context encoder processes the input and generates an encoded token sequence $\seq{\bmxi}{M_i}$ for next-step decoding.

\subsubsection{Transformer-based Interest-aware Decoder}
\label{subsubsec:mm_dec}
After obtaining the encoded interest tokens $\seq{\bmxi}{M_i}$, we design an interest-aware decoding process to model item-modality-interest relations for precise and comprehensive sequence representation. 
Specifically, we construct a decoder with $K_\text{d}$ Transformer decoder blocks. 
For the multi-head cross-attention module in each decoder block, we take $\seq{\bmxi}{M_i}$ as the key and value. 
We add item-wise positional embeddings $\bmP_i=[\seq{\bmp}{L_i}]$ to the item tokens $\bmX_i=[\seq{\bmx}{L_i}]$ and form the decoding queries. 
Let us denote $\bmY_i=[\seq{\bmy}{L_i}]$ as the decoded output \wrt $\bmX_i$, then the embedding (\ie the sequence representation) for the $i$-th user is produced by the output of the last query token, namely $\bmu_i=\bmy_{L_i}$. 

\subsection{Self-supervised Contrastive Pre-training}
\label{subsec:pretrain}
To pursue universal representation for multi-modal interaction sequences, we design two self-supervised contrastive learning (CL) tasks for pre-training the model. 

\subsubsection{Sequence-item Contrastive Learning}
\label{subsubsec:seq_item_cl}
To capture the correspondence between interaction sequences and candidate items, we take sequence-item CL as a pre-training task. The objective \wrt $\bmu_i$:
\begin{equation}
    \label{equ:SI-CL}
    \ell_i^{S-I}=-\log\frac
    {\exp({\langle\bmu_i,\bmv_{T_i+1}\rangle/\tau})}
    {\sum_{j=1}^B\exp({\langle\bmu_i,\bmv_{T_j+1}\rangle/\tau})}, 
\end{equation}
where $B$ denotes the size of mini-batch. $\tau>0$ us a temperature factor. 
The user-item matching scores are computed by \cref{equ:factorizable-ui-match-score}. 

\subsubsection{Sequence-sequence Contrastive Learning}
\label{subsubsec:seq_seq_cl}
Apart from sequence-item contrast, we leverage semantic invariance as a self-supervising signal to enhance the robustness of the sequence model. Given $\bmu_i,\bmu_i'$ as the representations of two augmented sequences (see \S\ref{subsubsec:feat_aug}) \wrt the $i$-th user, we define the sequence-sequence CL by
\begin{equation}
    \label{equ:SS-CL}
    \ell_i^{S-S}=-
    \log
    \frac
    {\exp({\langle\bmu_i,\bmu_i'\rangle/\tau})}
    {\sum_{j=1}^B
    \exp({\langle\bmu_i,\bmu_j\rangle/\tau}) + 
    \exp({\langle\bmu_i,\bmu_j'\rangle/\tau})
    }. 
\end{equation}

\subsubsection{Overall Objectives}
\label{subsubsec:overall_objs}
We sum up pre-train objectives as follows
\begin{gather}
    \label{equ:pre-train}
    \calL_\text{pre-train}=\frac1B\sum_{i=1}^B\zhongk{\ell_i^{S-I}+\lambda\cdot \ell_i^{S-S} + \frac{\gamma}{M_i^2} \sum_{m,m'=1}^{M_i}\langle\bmxi_m,\bmxi_{m'}\rangle}.
\end{gather}
The last term is an orthogonal regularization to diversify interest-aware decoded results. 
$\lambda, \gamma>0$ are the weights for different terms. 

\subsection{Efficient Fine-tuning}
\label{subsec:fine_tuning}
In order to transfer the recommendation knowledge to downstream domains and enhance their performance, we further study efficient fine-tuning based on the pre-trained sequence model. 
For the tuning objectives, we adopt sequence-item CL and orthogonal regularization to  directly optimize the next-item prediction, namely,
\begin{gather}
    \label{equ:fine-tune}
    \calL_\text{fine-tune}=\frac1B\sum_{i=1}^B\zhongk{\ell_i^{S-I}+ \frac{\gamma}{M_i^2} \sum_{m,m'=1}^{M_i}\langle\bmxi_m,\bmxi_{m'}\rangle}.
\end{gather}

As shown in the right block of \cref{fig:arc}, we disable the feature augmentation and the sequence-sequence contrast, so the sequence model only encodes once per user, which is computation-efficient. 
Besides, only the item ID embedding table (if included) and modality-specific adapters need to be tuned while other modules keep frozen, which is also parameter-efficient~\cite{UnifiedPETL}. 
Following ~\citet{UniSRec}, we support \emph{inductive} and \emph{transductive} transfer settings for \modelname{}. 

\subsubsection{Inductive Transfer}
\label{subsubsec:inductive_transfer}
Given a target domain with lots of cold items, making robust predictions is tricky for ID-based recommenders. 
We support an ID-agnostic solution to deal with extreme sparsity, where the model predicts next item for the $i$-th user by
\begin{gather}
    \text{Pr}_\text{ind}(I_{T_i+1}\mid \seq{I}{T_i})=\text{softmax}(\langle\bmu_i,\bmv_{T_i+1}\rangle).  
\end{gather}

\subsubsection{Transductive Transfer}
\label{subsubsec:transductive_transfer}
Given a target domain of mostly warm items, we pursue more precise recommendations by utilizing item IDs. 
Specifically, the model predicts next item for the $i$-th user by
\begin{gather}
    \text{Pr}_\text{trd}(I_{T_i+1}\mid \seq{I}{T_i})=\text{softmax}(\langle\bar{\bmu}_i,\bmv_{T_i+1}+\bmz_{T_i+1}\rangle),    
\end{gather}
where $\bmz_{T_i+1}$ is the ID embeddings of $I_{T_i+1}$. $\bar{\bmu}_i$ denotes the encoded sequence representation with item IDs (see the right block in \cref{fig:arc}).

\section{Experiments}
\label{sec:experiments}

\subsection{Research Questions}
\label{subsec:research_questions}
We evaluate the proposed method by conducting experiments on three datasets. We aim to answer the following research
questions:
\begin{itemize}
	\item[\textbf{RQ1:}] Compared with state-of-the-art sequential recommendation models using various types of information, can \modelname{} achieve competitive performance in downstream domains?
	\item[\textbf{RQ2:}] Is \modelname{} better than other modality-based baselines in utilizing different modalities? How much impact do multi-modal information and pre-training have on its efficacy? 
	\item[\textbf{RQ3:}] How do different designs contribute to \modelname{}'s efficacy?
\end{itemize}

\begin{table}[t] %
	\caption{Statistics of Pre-processed Datasets. ``Cover.'' denotes the image coverage among the item set. ``Avg. SL'' denotes the average length of interaction sequences.}
	\label{tab:dataset}
	\resizebox{\columnwidth}{!}{
	\begin{tabular}{l *{5}{r}}
		\toprule
		Datasets & \#Users & \#Items & \#Img. (Cover./\%) & \#Inters. & Avg. SL.\\
            \hline
		\midrule
		\emph{Pre-trained} & 1,361,408 & 446,975 & 94,151 (21.06\%) & 14,029,229 & 13.51 \\
		- Food   & 115,349 &  39,670 & 29,990 (75.60\%) & 1,027,413 &  8.91 \\
		- CDs    &  94,010 &  64,439 & 21,166 (32.85\%) & 1,118,563 & 12.64 \\
		- Kindle & 138,436 &  98,111 & 0 (0\%) & 2,204,596 & 15.93 \\
		- Movies & 281,700 &  59.203 & 8,675 (14.65\%) & 3,226,731 & 11.45 \\
		- Home   & 731,913 & 185,552 & 34,320 (18.50\%) & 6,451,926 &  8.82 \\
		\midrule
		Scientific  &  8,442 &  4,385 & 1,585 (36.15\%) &  59,427 & 7.04 \\
		Pantry      & 13,101 &  4,898 & 4,587 (93.65\%) & 126,962 & 9.69 \\
		Instruments & 24,962 &  9,964 & 6,289 (63.12\%) & 208,926 & 8.37 \\
		Arts        & 45,486 & 21,019 & 9,437 (44.90\%) & 395,150 & 8.69 \\
		Office      & 87,436 & 25,986 & 16,628 (63.99\%) & 684,837 & 7.84 \\ 
		\bottomrule
	\end{tabular}
	}
\end{table}

\subsection{Experimental Setup}
\label{subsec:setup}
\subsubsection{Datasets}
\label{subsubsec:dataset_setup}
We adopt 10 domains including ``\emph{Grocery and Gourmet Food}'', ``\emph{Home and Kitchen}'', ``\emph{CDs and Vinyl}'', ``\emph{Kindle Store}'', ``\emph{Movies and TV}'', ``\emph{Prime Pantry}'', ``\emph{Industrial and Scientific}'', ``\emph{Musical Instruments}'', ``\emph{Arts, Crafts and Sewing}'', and ``\emph{Office Products}'', from the standard benchmark dataset, \textbf{Amazon Review}~\cite{AmazonReview}. 
To provide extensive evaluations of the transferability, We divide the former 5 and the latter 5 into pre-training and downstream target domains, respectively. 
We follow \citet{UniSRec} to pre-process for interaction data and retrieve text information including \emph{title}, \emph{categories} and \emph{brand} from the metadata. 
To support multi-modal inputs, we further crawl item images according to the URLs in the metadata. 
The dataset statistics are reported in \cref{tab:dataset}. 
While texts are fully available in the metadata, there are many missing images in the dataset due to expired products or URLs. 
We still retain the items without images in our experiments to keep a fair comparison with previous works~\cite{UniSRec}.  
As we will show below, our \modelname{} can exhibit robust improvement on incomplete multi-modal datasets. 

\subsubsection{Metrics}
\label{subsubsec:metrics_setup}
Following previous works~\cite{UniSRec,S3-Rec}, we adopt two standard metrics, \ie \textbf{Recall} (\textbf{R@$K$}) and \textbf{NDCG} (\textbf{N@$K$}), to evaluate the retrieval performance. 
We set $K$ to 10 and 50 for showcases. 

\subsubsection{Baselines}
\label{subsubsec:baselines_setup}

We compare our method with 7 state-of-the-art sequential recommenders, including 
(\textbf{i}) 2 pure ID-based recommenders: SASRec~\cite{SASRec} and
BERT4Rec~\cite{BERT4Rec};
(\textbf{ii}) 1 attribute-aware recommender,  
S$^3$-Rec~\cite{S3-Rec}; 
(\textbf{iii}) 4 text-based recommenders, 
FDSA~\cite{FDSA}, 
ZESRec~\cite{ZESRec}, and
UniSRec~\cite{UniSRec}. 
We derive an extra text-feature-based variant from SASRec regarding its simplicity and effectiveness. 
For S$^3$-Rec, we inherit previous works~\cite{UniSRec} to embed the tags with textual features so as to support universal representation. 
We adopt both transductive and inductive variants of UniSRec for comparison. 
Therefore, we have 9 baselines in total.

\begin{table*}[t]
\centering
\caption{Comparisons on different target datasets. 
``T'' and ``V'' stands for text and visual features. 
``\emph{Improv}.'' denotes the relative improvement of \modelname{} to the best baselines. 
The best and second-best results are in bold and underlined, respectively.}
\setlength{\tabcolsep}{0.5em}{
 \resizebox{1.0\textwidth}{!}{
\begin{tabular}{lclccccccrlccccr}
\toprule
\multicolumn{2}{l}{\makecell{Input Type \& Model $\rightarrow$}} &  & \multicolumn{2}{c}{ID} & \multicolumn{3}{c}{T+ID} & T+V+ID & \multirow{2}{*}{\makecell{\emph{Improv.} \\ w/ ID}} &  & \multicolumn{3}{c}{T} & T+V & \multirow{2}{*}{\makecell{\emph{Improv.} \\ w/o ID}} \\
\cmidrule(l){1-2} \cmidrule(l){4-5} \cmidrule(l){6-8} \cmidrule(l){9-9} \cmidrule(l){12-14} \cmidrule(l){15-15}
Dataset & Metric &  & {\small SASRec} & {\footnotesize BERT4Rec} & FDSA & S$^3$-Rec & {\small UniSRec} & {\small\modelname{}} &  &  & {\small SASRec} & {\small ZESRec} & {\small UniSRec} & {\small\modelname{}} &  \\
\hline
\midrule
\multirow{4}{*}{Scientific} & R@10 &  & 0.1080 & 0.0488 & 0.0899 & 0.0525 & \underline{0.1235} & \textbf{0.1360} & \cellcolor[HTML]{E9EAFF}10.12\% &  & 0.0994 & 0.0851 & \underline{0.1188} & \textbf{0.1278} & \cellcolor[HTML]{EFFBEC}7.58\% \\
 & N@10 &  & 0.0553 & 0.0243 & 0.0580 & 0.0275 & \underline{0.0634} & \textbf{0.0753} & \cellcolor[HTML]{E9EAFF}18.77\% &  & 0.0561 & 0.0475 & \underline{0.0641} & \textbf{0.0658} & \cellcolor[HTML]{EFFBEC}2.65\% \\
 & R@50 &  & 0.2042 & 0.1185 & 0.1732 & 0.1418 & \textbf{0.2473} & \underline{0.2431} & \cellcolor[HTML]{E9EAFF}- &  & 0.2162 & 0.1746 & \textbf{0.2394} & \underline{0.2375} & \cellcolor[HTML]{EFFBEC}- \\
 & N@50 &  & 0.0760 & 0.0393 & 0.0759 & 0.0468 & \underline{0.0904} & \textbf{0.0983} & \cellcolor[HTML]{E9EAFF}8.74\% &  & 0.0815 & 0.0670 & \textbf{0.0903} & \underline{0.0893} & \cellcolor[HTML]{EFFBEC}- \\
\midrule
\multirow{4}{*}{Pantry} & R@10 &  & 0.0501 & 0.0308 & 0.0395 & 0.0444 & \underline{0.0693} & \textbf{0.0779} & \cellcolor[HTML]{E9EAFF}12.41\% &  & 0.0585 & 0.0454 & \underline{0.0636} & \textbf{0.0771} & \cellcolor[HTML]{EFFBEC}21.23\% \\
 & N@10 &  & 0.0218 & 0.0152 & 0.0209 & 0.0214 & \underline{0.0311} & \textbf{0.0365} & \cellcolor[HTML]{E9EAFF}17.36\% &  & 0.0285 & 0.0230 & \underline{0.0306} & \textbf{0.0345} & \cellcolor[HTML]{EFFBEC}12.75\% \\
 & R@50 &  & 0.1322 & 0.1030 & 0.1151 & 0.1315 & \underline{0.1827} & \textbf{0.1875} & \cellcolor[HTML]{E9EAFF}2.63\% &  & 0.1647 & 0.1141 & \underline{0.1658} & \textbf{0.1833} & \cellcolor[HTML]{EFFBEC}10.55\% \\
 & N@50 &  & 0.0394 & 0.0305 & 0.0370 & 0.0400 & \underline{0.0556} & \textbf{0.0598} & \cellcolor[HTML]{E9EAFF}7.55\% &  & 0.0523 & 0.0378 & \underline{0.0527} & \textbf{0.0571} & \cellcolor[HTML]{EFFBEC}8.35\% \\
\midrule
\multirow{4}{*}{Instruments} & R@10 &  & 0.1118 & 0.0813 & 0.1070 & 0.1056 & \underline{0.1267} & \textbf{0.1300} & \cellcolor[HTML]{E9EAFF}2.60\% &  & 0.1127 & 0.0783 & \underline{0.1189} & \textbf{0.1201} & \cellcolor[HTML]{EFFBEC}1.01\% \\
 & N@10 &  & 0.0612 & 0.0620 & \underline{0.0796} & 0.0713 & 0.0748 & \textbf{0.0843} & \cellcolor[HTML]{E9EAFF}5.90\% &  & 0.0661 & 0.0497 & \underline{0.0680} & \textbf{0.0771} & \cellcolor[HTML]{EFFBEC}13.38\% \\
 & R@50 &  & 0.2106 & 0.1454 & 0.1890 & 0.1927 & \textbf{0.2387} & \underline{0.2370} & \cellcolor[HTML]{E9EAFF}- &  & 0.2104 & 0.1387 & \textbf{0.2255} & \underline{0.2218} & \cellcolor[HTML]{EFFBEC}- \\
 & N@50 &  & 0.0826 & 0.0756 & 0.0972 & 0.0901 & \underline{0.0991} & \textbf{0.1071} & \cellcolor[HTML]{E9EAFF}8.07\% &  & 0.0873 & 0.0627 & \underline{0.0912} & \textbf{0.0988} & \cellcolor[HTML]{EFFBEC}8.33\% \\
\midrule
\multirow{4}{*}{Arts} & R@10 &  & 0.1108 & 0.0722 & 0.1002 & 0.1003 & \underline{0.1239} & \textbf{0.1314} & \cellcolor[HTML]{E9EAFF}6.05\% &  & 0.0977 & 0.0664 & \underline{0.1066} & \textbf{0.1119} & \cellcolor[HTML]{EFFBEC}4.97\% \\
 & N@10 &  & 0.0587 & 0.0479 & \underline{0.0714} & 0.0601 & 0.0712 & \textbf{0.0767} & \cellcolor[HTML]{E9EAFF}7.42\% &  & 0.0562 & 0.0375 & \underline{0.0586} & \textbf{0.0625} & \cellcolor[HTML]{EFFBEC}6.66\% \\
 & R@50 &  & 0.2030 & 0.1367 & 0.1779 & 0.1888 & \underline{0.2347} & \textbf{0.2410} & \cellcolor[HTML]{E9EAFF}2.68\% &  & 0.1916 & 0.1323 & \underline{0.2049} & \textbf{0.2100} & \cellcolor[HTML]{EFFBEC}2.49\% \\
 & N@50 &  & 0.0788 & 0.0619 & 0.0883 & 0.0793 & \underline{0.0955} & \textbf{0.1002} & \cellcolor[HTML]{E9EAFF}4.92\% &  & 0.0766 & 0.0518 & \underline{0.0799} & \textbf{0.0836} & \cellcolor[HTML]{EFFBEC}4.63\% \\
\midrule
\multirow{4}{*}{Office} & R@10 &  & 0.1056 & 0.0825 & 0.1118 & 0.1030 & \textbf{0.1280} & \underline{0.1275} & \cellcolor[HTML]{E9EAFF}- &  & 0.0929 & 0.0641 & \underline{0.1013} & \textbf{0.1038} & \cellcolor[HTML]{EFFBEC}2.47\% \\
 & N@10 &  & 0.0710 & 0.0634 & \textbf{0.0868} & 0.0653 & 0.0831 & \underline{0.0856} & \cellcolor[HTML]{E9EAFF}- &  & 0.0582 & 0.0391 & \underline{0.0619} & \textbf{0.0666} & \cellcolor[HTML]{EFFBEC}7.59\% \\
 & R@50 &  & 0.1627 & 0.1227 & 0.1665 & 0.1613 & \textbf{0.2016} & \underline{0.2005} & \cellcolor[HTML]{E9EAFF}- &  & 0.1580 & 0.1113 & \textbf{0.1702} & \underline{0.1701} & \cellcolor[HTML]{EFFBEC}- \\
 & N@50 &  & 0.0835 & 0.0721 & 0.0987 & 0.0780 & \underline{0.0991} & \textbf{0.1012} & \cellcolor[HTML]{E9EAFF}2.12\% &  & 0.0723 & 0.0493 & \underline{0.0769} & \textbf{0.0808} & \cellcolor[HTML]{EFFBEC}5.07\% \\
\hline
\bottomrule
\end{tabular}
}
}
\label{tab:main_results}
\end{table*}

\subsubsection{Implementation Details}
\label{subsubsec:baselines_setup}
We take Adam as the common optimizer in our comparison, and we carefully choose suitable learning rates from $\{3e^{-4}, 1e^{-3}, 3e^{-3}, 1e^{-2}\}$ for different baselines. 
We take 100 epochs for pre-training. 
And for fine-tuning, we implement early stopping with a patience of 10 to prevent over-fitting. 
We adopt CLIP-B/32~\cite{CLIP,HuggingHash} as the basic feature encoder to extract 
\texttt{[CLS]} features for each text and image. 
The modality adapters then transform the features into the 300-d latent space for sequence representation. 
In the MID module (\S\ref{subsec:mid_module}), the number of interest prototypes, $K_\text{c}$, is set to 2\% of the item amount during pre-training, and is set to 50\% of the item amount during fine-tuning. 
To keep a consistent parameter scale with baselines, we build our model with 1 Transformer encoder block and 1 decoder block, \ie $K_\text{e}=K_\text{d}=1$. 
The loss weights in \cref{equ:pre-train,equ:fine-tune} are set $\lambda=1e^{-3}$ and $\gamma=1e^{-4}$, respectively. 
The temperature factor for contrastive learning objectives, \eg \cref{equ:SI-CL,equ:SS-CL}, is set $\tau=0.07$. 
The concentration factor $\alpha$ in \S\ref{subsubsec:candidate_item_enc} is set to a learnable variable. 
The data and code have been released on \url{https://github.com/gimpong/MM23-MISSRec}.

\begin{figure}[t]
    \centering
    \includegraphics[width=\columnwidth]{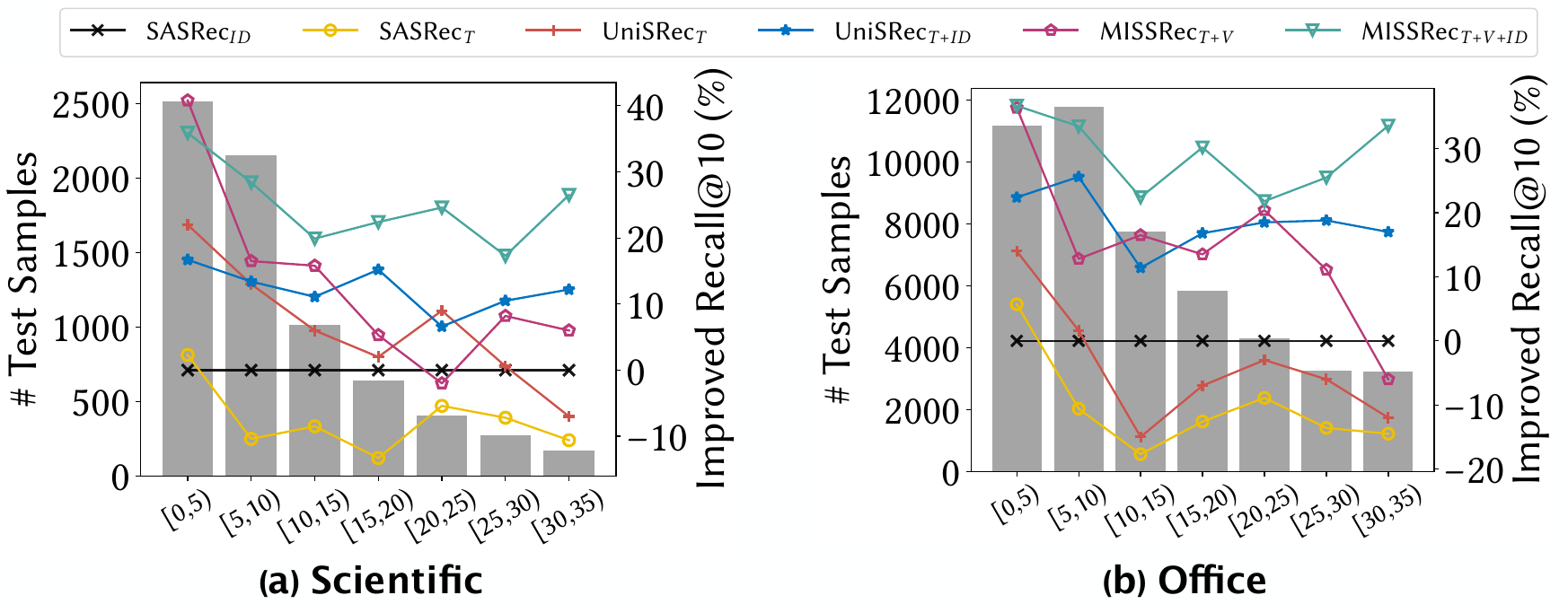}
    \caption{Performance comparison on long-tailed items. 
    The histogram represents the number of test samples (\ie interacted items) of different popularities. 
    The line chart shows the relative improvement to SASRec$_\text{ID}$ on Recall@10.}
\label{fig:cold_start_analysis}
\end{figure}

\subsection{Comparasion with State-of-the-arts (RQ1)}
\label{subsec:sota}
\subsubsection{Overall Performance}
\label{subsubsec:ovearll_sota}
The overall comparison results on five different downstream datasets are shown in \cref{tab:main_results}, from which we obtain three findings. 
(\textbf{i}) Text features can be exploited as an effective alternative or supplement to ID features, and pre-training can further enhance their efficacy. 
Comparing the SASRec variants with different types of input, the text-based variant achieves competitive performance and outperforms the ID-based variant on ``Pantry'' and ``Instruments'' datasets. 
Besides, text-enhanced approaches FDSA and S$^3$-Rec outperform ID-based BERT4Rec by large margins. 
By adopting pre-training, UniSRec achieves further improvement. 
(\textbf{ii}) ID information is still important for personalization. 
Transductive methods generally perform better than inductive ones. 
(\textbf{iii}) Under both transductive and inductive settings, our \modelname{} outperforms state-of-the-art baselines in most cases, and the margins are considerable. 
We attribute the performance gain to three factors. 
First, multi-modal information reflects more precise and comprehensive user preferences, benefiting personalization. 
Second, the synergy of different designs in \modelname{} contributes to the efficacy of multi-modal sequence representation. 
Third, pre-training further strengthened the above advantages.

\subsubsection{Performance \wrt Different Item Frequencies}
\label{subsubsec:frequency_sota}
In this sub-section, we select the best performers from \cref{tab:main_results} and compare their efficacy \wrt long-tailed items. 
As shown in \cref{fig:cold_start_analysis}, all modality-based methods demonstrate effectiveness to ID-based SASRec on the least frequent items, while the performance gains begin to vanish and even turn to degradation as the items warm up. 
Compared with other baselines, \modelname{} with multi-modal learning shows the best few-short performance and also keeps robust improvement with more frequent items, suggesting its feasibility in addressing long-tailed and cold-start recommendation issues.

\subsection{Multi-modal \& Pre-training Analyses (RQ2)}
\label{subsec:mm_pretrain_analyses}

\begin{table}[t] %
	\caption{Model comparison \wrt various input types on the full-modality data subset. We can see the superior capability of \modelname{} in leveraging modality features.}
        \setlength{\tabcolsep}{0.3em}{
	\resizebox{\columnwidth}{!}{
        \begin{tabular}{cccccccccccc}
        \toprule
        &  &  & \multicolumn{4}{c}{Scientific} &  & \multicolumn{4}{c}{Office} \\
        \cmidrule(l){4-7} \cmidrule{9-12}
        Input Type & Variant &  & R@10 & R@50 & N@10 & N@50 &  & R@10 & R@50 & N@10 & N@50 \\
        \hline
        \midrule
        \multirow{2}{*}{ID+T} & UniSRec &  & 0.1676 & 0.3200 & 0.0872 & 0.1208 &  & 0.1376 & 0.2152 & 0.0937 & 0.1104 \\
        & \modelname{} &  & \textbf{0.1697} & \textbf{0.3249} & \textbf{0.0890} & \textbf{0.1227} &  & \textbf{0.1398} & \textbf{0.2195} & \textbf{0.0953} & \textbf{0.1122} \\
        \midrule
        \multirow{2}{*}{ID+V} & UniSRec &  & 0.1631 & 0.3112 & 0.0857 & 0.1188 &  & 0.1324 & 0.2071 & 0.0921 & 0.1086 \\
        & \modelname{} &  & \textbf{0.1657} & \textbf{0.3165} & \textbf{0.0869} & \textbf{0.1198} &  & \textbf{0.1375} & \textbf{0.2152} & \textbf{0.0946} & \textbf{0.1113} \\
        \midrule
        \multirow{2}{*}{ID+T+V} & UniSRec &  & 0.1680 & 0.3209 & 0.0865 & 0.1194 &  & 0.1407 & 0.2203 & 0.0957 & 0.1133 \\
        & \modelname{} &  & \textbf{0.1711} & \textbf{0.3265} & \textbf{0.0890} & \textbf{0.1227} &  & \textbf{0.1421} & \textbf{0.2223} & \textbf{0.0966} & \textbf{0.1138} \\
        \midrule
        \multirow{2}{*}{T} & UniSRec &  & 0.1602 & 0.3202 & 0.0806 & 0.1148 &  & 0.1183 & 0.1893 & 0.0773 & 0.0926 \\
        & \modelname{} &  & \textbf{0.1619} & \textbf{0.3239} & \textbf{0.0814} & \textbf{0.1165} &  & \textbf{0.1186} & \textbf{0.1899} & \textbf{0.0787} & \textbf{0.0943} \\
        \midrule
        \multirow{2}{*}{V} & UniSRec &  & 0.1568 & 0.3135 & 0.0782 & 0.1118 &  & 0.1140 & 0.1823 & 0.0763 & 0.0910 \\
        & \modelname{} &  & \textbf{0.1575} & \textbf{0.3144} & \textbf{0.0789} & \textbf{0.1118} &  & \textbf{0.1152} & \textbf{0.1848} & \textbf{0.0771} & \textbf{0.0923} \\
        \midrule
        \multirow{2}{*}{T+V} & UniSRec &  & 0.1609 & 0.3216 & 0.0808 & 0.1141 &  & 0.1205 & 0.1929 & 0.0792 & 0.0945 \\
        & \modelname{} &  & \textbf{0.1633} & \textbf{0.3265} & \textbf{0.0812} & \textbf{0.1160} &  & \textbf{0.1209} & \textbf{0.1936} & \textbf{0.0798} & \textbf{0.0953} \\
        \bottomrule
        \end{tabular}
	}
        }
	\label{tab:modality_analysis}
\end{table}

\subsubsection{Comparison with Different Input Modalities}
\label{subsubsec:compare_with_diff_modality}
In this sub-section, we investigate the capacities of different modality-based sequence models in leveraging multi-modal information. 
Specifically, we filter out the items without images in ``Scientific'' and ``Office'' datasets. 
Then we conduct experiments on the filtered subsets to enable horizontal comparison among different settings, \eg visual-only \emph{vs} text-only. 
To ensure objective evaluation, we adopt the same CLIP-B/32 feature extractors for all comparison methods. 
We report the results in \cref{tab:modality_analysis}. 
Interestingly, although utilizing visual modality alone does not lead to satisfactory results, the multi-modal synergy creates an effect of ``1+1>2'', which highlights the significance of multi-modal sequential modeling. 
Besides, \modelname{} demonstrates robust and competitive performance under various settings.

\begin{table}[t] %
	\caption{Analysis of the effect of pre-training.}
        \setlength{\tabcolsep}{0.2em}{
	\resizebox{\columnwidth}{!}{
        \begin{tabular}{cclcccclcccc}
        \toprule
        \multicolumn{2}{c}{\underline{Model: \modelname{}}} &  & \multicolumn{4}{c}{Scientific} &  & \multicolumn{4}{c}{Office} \\
        \cmidrule(l){4-7} \cmidrule(l){9-12}
        w/ ID? & Pre-train? &  & R@10 & R@50 & N@10 & N@50 &  & R@10 & R@50 & N@10 & N@50 \\
        \hline
        \midrule
        \multirow{2}{*}{\dui} & \dui &  & 0.1360 & 0.2431 & 0.0753 & 0.0983 &  & 0.1275 & 0.2005 & 0.0856 & 0.1012 \\
        & \cuo &  & 0.1282 & 0.2376 & 0.0711 & 0.0946 &  & 0.1269 & 0.2001 & 0.0848 & 0.1005 \\
        \multicolumn{2}{c}{\cellcolor[HTML]{E9EAFF}\emph{Improv.} w/ ID} && \cellcolor[HTML]{E9EAFF}6.08\% & \cellcolor[HTML]{E9EAFF}2.31\% & \cellcolor[HTML]{E9EAFF}5.91\% & \cellcolor[HTML]{E9EAFF}3.91\% &  & \cellcolor[HTML]{E9EAFF}0.47\% & \cellcolor[HTML]{E9EAFF}0.20\% & \cellcolor[HTML]{E9EAFF}0.94\% & \cellcolor[HTML]{E9EAFF}0.70\% \\
        \midrule
        \multirow{2}{*}{\cuo} & \dui &  & 0.1278 & 0.2375 & 0.0658 & 0.0893 &  & 0.1038 & 0.1701 & 0.0666 & 0.0808 \\
        & \cuo &  & 0.1269 & 0.2354 & 0.0659 & 0.0891 &  & 0.1072 & 0.1726 & 0.0694 & 0.0834 \\
        \multicolumn{2}{c}{\cellcolor[HTML]{EFFBEC}\emph{Improv.} w/o ID} && \cellcolor[HTML]{EFFBEC}0.71\% & \cellcolor[HTML]{EFFBEC}0.89\% & \cellcolor[HTML]{EFFBEC}-0.15\% & \cellcolor[HTML]{EFFBEC}0.22\% & \cellcolor[HTML]{EFFBEC} & \cellcolor[HTML]{EFFBEC}-3.17\% & \cellcolor[HTML]{EFFBEC}-1.45\% & \cellcolor[HTML]{EFFBEC}-4.03\% & \cellcolor[HTML]{EFFBEC}-3.12\% \\
        \bottomrule
        \end{tabular}
	}
        }
	\label{tab:pretrain_analysis}
\end{table}

\subsubsection{Pre-training Analysis}
\label{subsubsec:pretraining_effect}
Here we examine the effectiveness of pre-training to \modelname{}. 
As shown in \cref{tab:pretrain_analysis}, we can see that pre-training enhances recommendation performance in target domains under transductive settings. 
However, we observe the negative transfer under inductive settings, especially on the larger dataset, Office. 
A possible reason for the phenomenon is 
that the model can not adapt well to the target data with a large domain gap accompanied by quite limited tunable parameters (\ie adapters).  

\subsection{Model Analyses (RQ3)}
\label{subsec:model_analyses}
To investigate the efficacy of different components, we construct 5 variants of \modelname{}$_\text{T+V+ID}$ for comparison. 
Specifically, as presented by the leftmost column in \cref{tab:ablation_study}, variant (1) removes the sequence-sequence contrastive task (\S\ref{subsubsec:seq_seq_cl}) from pre-training. Variant (2) removes the modality-specific adapters (\S\ref{subsubsec:adapter}) so that the sequence model directly consumes 512-d CLIP features. 
Variant (3) replaces the interest-aware decoder (\S\ref{subsubsec:mm_dec}) layer with another encoder layer and builds a 2-layer contextual encoder that only processes interest tokens. 
The sequence embedding is produced by mean-pooling the output token sequence. 
Variant (4) only keeps the decoding structure and builds a 2-layer decoder-only model. 
It also removes the cross-attention module at each decoder layer due to the absence of encoder output. 
Variant (5) removes the orthogonal regularization in learning objectives (\cref{equ:pre-train,equ:fine-tune}).

\subsubsection{Effects of Modality Adapters}
\label{subsubsec:analyze_adapters}
Comparing variants (2) with (0), we can see significant performance decays, owing to the semantic gap between general multi-modal features and user interests.
As a result, multi-modal features themselves as item representations could not provide sufficient personalization for recommendation, and modality adapters are required. 
On the other hand, although end-to-end optimizing feature extractors with the recommender also helps to mitigate the gap, as revealed by \cite{MoRec,TransRec}, it will consume much more computation resources. 
Anyway, parameter-efficient tuning with modality adapters can be preferable. 

\subsubsection{Effects of Learning Objectives}
\label{subsubsec:analyze_loss}
Comparing (1) and (0), we learn that $\ell_i^\text{S-S}$ in the pre-training stage can enhance the representation robustness, as it helps to improve downstream recommendation. 
Comparing (5) and (0), we find that the orthogonal regularization leads to slightly better results on the Scientific dataset but shows a negative impact on the Office dataset. 
We can infer that the orthogonal regularization is sensitive to the domain or the weighting factor, which requires careful choosing according to specific scenarios. 

\subsubsection{Effects of Interest-aware Decoding and Contextual Encoding}
\label{subsubsec:analyze_enc_dec}
We can learn that variant (3) exhibits poor performance among the variants, suggesting the necessity of the collaboration between context encoding and interest-aware decoding. 
Besides, by comparing variants (4) and (0), variant (4) without accessing encoded interest information also shows performance degradation, which justifies the efficacy of interest mining and contextual modeling. 

\begin{table}[t] %
	\caption{Ablation study with \modelname{}.}
        \setlength{\tabcolsep}{0.2em}{
	\resizebox{\columnwidth}{!}{
        \begin{tabular}{llcccclcccc}
        \toprule
         &  & \multicolumn{4}{c}{Scientific} &  & \multicolumn{4}{c}{Office} \\
        \cmidrule(l){3-6} \cmidrule(l){8-11}
        Variant &  & R@10 & R@50 & N@10 & N@50 &  & R@10 & R@50 & N@10 & N@50 \\
        \hline
        \midrule
        (0) \modelname{} && \textbf{0.1360} & \textbf{0.2431} & \textbf{0.0753} & \textbf{0.0983} && \underline{0.1275} & \underline{0.2005} & \underline{0.0856} & \underline{0.1012} \\
        \cmidrule(l){1-1}
        (1) \phantom{-} w/o $\ell^{S-S}_i$ in Eq.(\ref{equ:pre-train}) && 0.1337 & 0.2390 & 0.0710 & 0.0926 && 0.1273 & 0.2000 & 0.0847 & 0.1002 \\
        (2) \phantom{-} w/o Modality Adapters && 0.1058 & 0.1896 & 0.0582 & 0.0759 && 0.1029 & 0.1617 & 0.0708 & 0.0838 \\
        (3) \phantom{-} 2-Layer Encoder Only && 0.1220 & 0.2186 & 0.0680 & 0.0890 && 0.1175 & 0.1849 & 0.0823 & 0.0974 \\
        (4) \phantom{-} 2-Layer Decoder Only && 0.1172 & 0.2092 & 0.0647 & 0.0844 && 0.1003 & 0.1575 & 0.0696 & 0.0821 \\
        (5) \phantom{-} w/o Oth. Reg. in Eqs.(\ref{equ:pre-train}-\ref{equ:fine-tune}) && \underline{0.1355} & \underline{0.2403} & \underline{0.0745} & \underline{0.0974} && \textbf{0.1286} & \textbf{0.2019} & \textbf{0.0873} & \textbf{0.1034} \\
        \bottomrule
        \end{tabular}
	}
        }
	\label{tab:ablation_study}
\end{table}
\section{Conclusions}
\label{sec:conclusion}

This paper addresses the limitations of existing sequential recommendation models that rely heavily on ID features by exploring the potential of multi-modal information. 
We propose \modelname{}, a novel multi-modal pre-training and transfer learning framework for sequential recommendation, which effectively tackles the cold-start problem and allows for efficient domain adaptation. 
By utilizing a transformer-based contextual encoder, an interest-aware decoder, a lightweight dynamic fusion module, 
\modelname{} demonstrates improved performance and generalizability compared to existing methods. 
Extensive experiments also showcase the compatibility and robustness of \modelname{} in handling incomplete or missing modalities, reinforcing its pragmatic value for real-world scenarios. 
More importantly, our paper suggests a promising direction for future research in leveraging multi-modal information for sequential recommendation. 
We hope \modelname{} can provide some inspiration to further explorations.

\begin{acks}
We want to thank the anonymous reviewers and the meta-reviewer for their valuable comments and suggestions. 
This research is supported by National Natural Science Foundation of China (Grant No.62171248 and Grant No.62276154), the Natural Science Foundation of Guangdong Province (Grant No.  2023A1515012914), Shenzhen Science and Technology Program (JCYJ20220818101012025), Basic Research Fund of Shenzhen City (Grant No. JCYJ20210324120012033 and JSGG20210802154402007), the Major Key Project of PCL for Experiments and Applications (PCL2021A07, and PCL2021A06), and Overseas Cooperation Research Fund of Tsinghua Shenzhen International
Graduate School (HW2021008).
We gratefully acknowledge the support of MindSpore\footnote{https://www.mindspore.cn}, which is a new deep learning framework used for this research.
\end{acks}

\bibliographystyle{ACM-Reference-Format}
\balance
\bibliography{main}

\end{document}